\documentclass{emulateapj}
\shorttitle{Cleaning up Eta Carinae}
\shortauthors{Smith et al.}
\begin{document}

\title{CLEANING UP ETA CARINAE: DETECTION OF AMMONIA IN THE HOMUNCULUS}

\author{Nathan Smith\altaffilmark{1,2}, Kate J.\
  Brooks\altaffilmark{3}, B\"{a}rbel S.\ Koribalski\altaffilmark{3},
  and John Bally\altaffilmark{1} }

\altaffiltext{1}{Center for Astrophysics and Space Astronomy,
University of Colorado, 389 UCB, Boulder, CO 80309}

\altaffiltext{2}{Hubble Fellow; nathans@casa.colorado.edu}

\altaffiltext{3}{Australia Telescope National Facility, CSIRO, PO Box
76, Epping, NSW 1710, Australia}

\begin{abstract}

We report the first detection of ammonia in the Homunculus nebula
around $\eta$ Carinae, which is also the first detection of emission
from a polyatomic molecule in this or any other luminous blue variable
(LBV) nebula.  Observations of the NH$_3$ (J,K)=(3,3) inversion
transition made with the Australia Telescope Compact Array reveal
emission at locations where infrared H$_2$ emission had been detected
previously, near the strongest dust emission in the core of the
Homunculus.  We also detect ammonia emission from the so-called
``strontium filament'' in the equatorial disk. The presence of NH$_3$
around $\eta$~Car hints that molecular shells around some Wolf-Rayet
stars could have originated in prior LBV eruptions, rather than in
cool red supergiant winds or the ambient interstellar medium.
Combined with the lack of any CO detection, NH$_3$ seems to suggest
that the Homunculus is nitrogen rich like the ionized ejecta around
$\eta$~Car.  It also indicates that the Homunculus is a unique
laboratory in which to study unusual molecule and dust chemistry, as
well as their rapid formation in a nitrogen-rich environment around a
hot star. We encourage future observations of other transitions like
NH$_3$ (1,1) and (2,2), related molecules like N$_2$H$^+$, and renewed
attempts to detect CO.

\end{abstract}


%%%%%%%%%%%%%%%%%%%%%%%%%%%%%%%%%%%%%%%%%%%%%%%%%%%%%%%%%%%%%%%%
\section{INTRODUCTION}

So far, $\eta$~Carinae is the only luminous blue variable (LBV) known
to be surrounded by dense molecular gas in its own ejecta nebula
(Smith 2006, 2002a; Smith \& Davidson 2001).  Nearly all other LBV
nebulae show bright infrared [Fe~{\sc ii}] emission from dense
partially-ionized gas (Smith 2002b; Smith \& Hartigan 2006) and many
have detectable dust shells (e.g., Clark et al.\ 2003).  This owes to
the extreme youth, high density, and extremely high mass (more than 10
M$_{\odot}$) of the bipolar Homunculus Nebula (Smith et al.\ 2003).
Thus, it is a unique laboratory in which to study the behavior of
molecules bathed in FUV radiation (e.g., Ferland et al.\ 2005).

While red supergiant (RSG) winds are familiar sites for molecule
detection, molecular gas in the ejecta of luminous hot stars is rare.
Some Wolf-Rayet (WR) stars are surrounded by molecular gas, but its
origin is not always clear.  The molecular gas traced by H$_2$ and CO
in NGC~2359, NGC~6888, and the nebula around WR~134 is thought to be
swept-up from the surrounding interstellar medium (St-Louis et al.\
1998; Cappa et al.\ 2001; Rizzo et al.\ 2003).  Several polyatomic
molecules have been detected in this swept-up material around NGC2359,
including ammonia (Rizzo et al.\ 2001).  The molecular gas surrounding
WR~16, WR~18 (NGC~3199) and WR~75 (RCW~104), on the other hand, is
thought to have formed from stellar ejecta in a previous RSG
evolutionary phase (Marston et al.\ 1999; Marston 2001; Welzmiller et
al.\ 1998).  In particular, HCN, HCO$^+$, CN, and HNC have been
detected in the NGC~3199 ring nebula around WR~18 (Marston 2001).  The
molecular gas in the Homunculus, in stark contrast, is known to have
formed out of ejecta from the 19th century outburst. If a WR
evolutionary phase will follow the current LBV phase of $\eta$ Car,
then its molecular gas may be relevant to the molecular shells around
WR stars -- i.e., it hints that they may have formed in giant LBV
eruptions like $\eta$ Car's 19th century outburst, not just in RSG
winds.  Several M$_{\odot}$ of material is present in these molecular
WR nebulae as well as the Homunculus (Marston et al.\ 1999; Welzmiller
et al.\ 1998; Smith et al.\ 2003).

Surrounding the Homunculus are the shock-heated ``outer ejecta'',
which are ionized and well-suited for nebular abundance studies,
revealing nitrogen enriched and C+O depleted material with n(N)/n(O)
of roughly 20 (Davidson et al.\ 1982; Dufour et al.\ 1997; Smith \&
Morse 2004).  The chemical composition of the Humongulus itself,
however, is unknown because it is mostly neutral or molecular.  No CO
has been detected (e.g., Cox \& Bronfman 1995), but its near-IR
emission lines of H$_2$ are bright (Smith 2002a, 2004, 2006).
Therefore, if it is composed of nitrogen-rich CNO ash like the other
ejecta around $\eta$ Car, one might expect to find emission from
ammonia.

We searched for emission from the NH$_3$ (J,K)=(3,3) inversion
transition instead of the (1,1) or (2,2) transitions for two reasons.
First, H$_2$ emission in the Homunculus -- while narrow at any single
position -- is spread over velocities of $\pm$500 km s$^{-1}$ (Smith
2002a, 2006).  This would blend the (1,1) and (2,2) transitions, as
they are separated by only 0.029 GHz or 367 km s$^{-1}$.  Second, the
energy above ground for the (3,3) transition (roughly 130 K) is
similar to the observed dust temperature of 140 K in the outer H$_2$
skin (Smith 2006; Smith et al.\ 2003).

%% Table 1
\begin{deluxetable}{lcccc}\tabletypesize{\scriptsize}
\tablecaption{Observations and Continuum Fluxes}\tablewidth{0pt}
\tablehead{
 \colhead{Date} &\colhead{Array}             &\colhead{Observing} 
                &\colhead{Cont.\ S$_{\nu}$}  &\colhead{Beam} \\
 \colhead{\ }   &\colhead{Config.}           &\colhead{time} 
                &\colhead{(Jy)}              &\colhead{Size}  }
\startdata
2004 Oct 30	&1.5C	&12 h	&4.01	&0$\farcs$5$\times$0$\farcs$4	\\
2005 Mar 29	&6A	&12 h	&2.47	&0$\farcs$4$\times$0$\farcs$3	\\
2005 Jul 21	&H75	&6 h	&7.39	&21\arcsec$\times$17\arcsec	\\
%(all three)	&...	&x$\farcs$x$\times$x$\farcs$x	&...	&	\\ \hline
\enddata
%\tablecomments{}
%\tablerefs{}
\end{deluxetable}

%%%%%%%%%%%%%%%%%%%%%%%%%%%%%%%%%%%%%%%%%%%%%%%%%%%%%%%%%%%%%%%%
\section{OBSERVATIONS}

We obtained observations of NH$_3$(3,3) emission toward $\eta$~Carinae
using three configurations of the Australia Telescope Compact Array
(ATCA) between 2004 October and 2005 June (Table 1). A bandwidth of
128 MHz was centered on 23.870 GHz, close to the rest frequency of the
ammonia NH$_3$(3,3) transition ($\nu_{\rm rest}$ = 23.870129 GHz), and
divided into 64 channels, giving an effective 2-channel velocity
resolution of $\sim$50 km s$^{-1}$ and velocity bandwidth of
$\sim$1200 km s$^{-1}$.  The flux was calibrated by observing
PKS\,B1934--638 and adopting a value of 0.77 Jy. We estimate the final
flux calibration to be better than 20\%. The bandpass and phase
calibrators were PKS\,1921-293 and PKS B1045--62, respectively. Data
were edited and calibrated using the {\sc miriad} software package
according to standard procedures.

For each individual dataset, continuum was subtracted in the {\em uv}
plane. The resultant line data were then combined and
Fourier-transformed into a data cube ($\alpha$, $\delta$, $v$). Each
channel image was deconvolved using the {\sc clean} algorithm and
restored with synthesized beams of either 3$\farcs$2$\times$2$\farcs$9
(Fig.\ 1; natural weighting) or 1$\farcs$0$\times$0$\farcs$7 (Fig.\ 2;
uniform weighting). The rms noise taken from an emission-free channel
is $\sim$0.4 mJy\,beam$^{-1}$ per 25 km s$^{-1}$.

In the case of the continuum emission, each of the three datasets were
Fourier-transformed separately and the images were deconvolved using
the {\sc clean} algorithm. The restoring synthesized beams as well as
the values for the measured integrated total emission (within the
10$\sigma$ contour level) are listed in Table 1. The fluxes vary
depending on the different array configurations as well as the
intrinsic millimeter variability of $\eta$ Car during its 5.5 yr cycle
(e.g., White et al.\ 2005).  From the SED of $\eta$ Car, we expect a
continuum flux of 2--3 Jy at 12mm (Brooks et al.\ 2005), which is
roughly consistent with the observed values listed in Table~1.  We do
not expect the ammonia emission or absorption to vary as much as the
continuum emission during the 5.5 yr cycle, because the molecular gas
is shielded from the weak changing Lyman continuum, and no significant
change in the near-IR H$_2$ emission is seen (Smith 2006).

% FIGURE 1 ---------- 
\begin{figure}
\epsscale{0.95}
\plotone{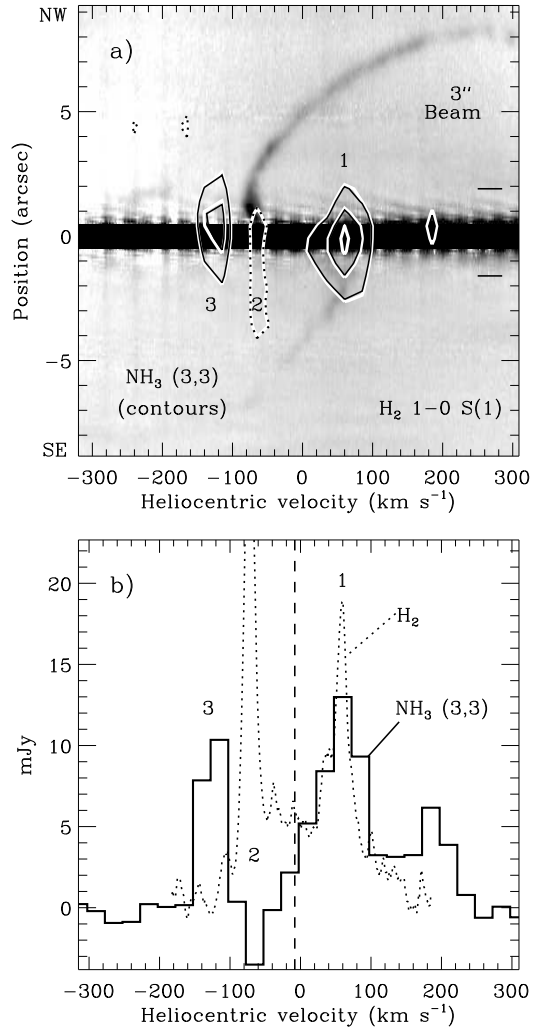}
\caption{(a) NH$_3$ (3,3) contours superposed on a position-velocity
  diagram of H$_2$~1$-$0~S(1) at 2.1218~$\micron$ (from Smith 2004).
  The ammonia emission was integrated over a $\sim$4\arcsec\ wide
  segment of the NH$_3$ (3,3) data cube, rotated counter-clockwise by
  45\arcdeg\ on the sky to be roughly aligned with the H$_2$ slit.
  Solid contours are drawn at 4 (10$\sigma$), 6.5, and 9 mJy/beam.
  The dashed contour is absorption at --1.3 mJy/beam
  ($\sim$3$\sigma$).  (b) The solid histogram is the
  continuum-subtracted spectrum of the NH$_3$ (3,3) intensity
  integrated over roughly 1 beam.  The dashed spectrum is H$_2$
  emission at $\pm$2$\farcs$3, and the vertical dashed line shows the
  systemic velocity of --8.1 km s$^{-1}$ (Smith 2004).}
\end{figure}

% FIGURE 2 ---------- 
\begin{figure}
\epsscale{0.98}
\plotone{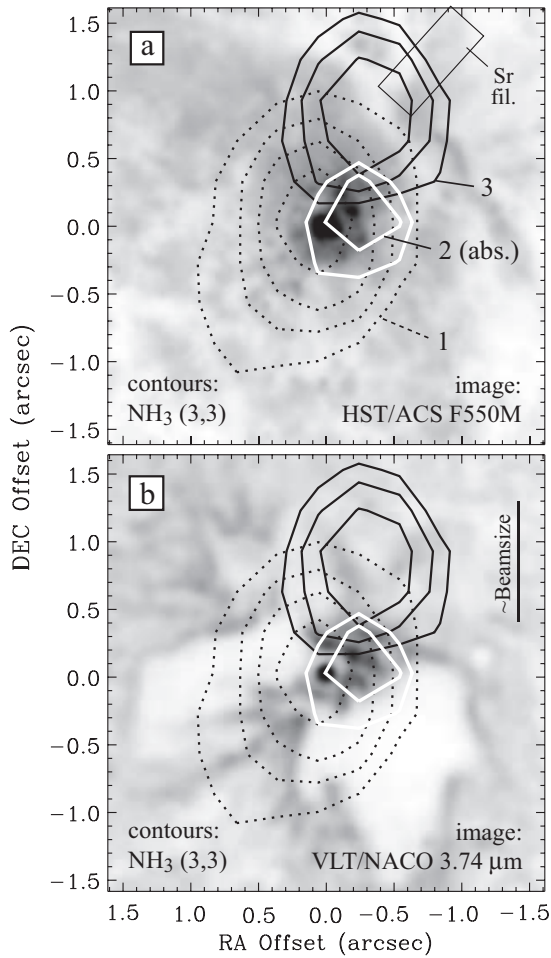}
\caption{The spatial distribution of ammonia compared to optical and
  IR emission.  (a) NH$_3$ (3,3) contours superposed on an {\it
  HST}/ACS image in visual continuum light (from Smith et al.\ 2004).
  Components 1, 2, and 3 are the same as in Figure 1 (see text).
  Contour levels are 4, 7, 10, and 13 mJy beam$^{-1}$ for component 1,
  --2 and --2.5 mJy beam$^{-1}$ for component 2 (absorption), and 3,
  5, and 7 mJy beam$^{-1}$ for component 3.  The ammonia data used
  here have been synthesized with a somewhat smaller $\sim$0$\farcs$9
  circular beam than in Fig.\ 1, favoring spatial resolution at the
  expense of sensitivity.  (b) The same contours superposed on a 3.74
  $\micron$ image (from Chesneau et al.\ 2005).  The rectangle in
  Panel $a$ denotes the approximate visual-wavelength extent of the
  ``strontium filament'', from Hartman et al.\ (2004).  The NH$_3$
  data and the images were spatially aligned by assuming that the
  centroid of 12mm continuum emission is the position of the star.}
\end{figure}

%%%%%%%%%%%%%%%%%%%%%%%%%%%%%%%%%%%%%%%%%%%%%%%%%%%%%%%%%%%%%%%%%%%%%%%
\section{RESULTS AND INTERPRETATION}

Figure 1 shows a position-velocity diagram and spectral tracings
comparing NH$_3$~(3,3) to H$_2$ $v$=1--0 S(1) from Smith (2004).  Both
lines have been shifted to a heliocentric velocity frame ($v_{\rm
hel}$=$v_{\rm LSR}$+11.6 km s$^{-1}$).  Figure 2 shows the spatial
distribution of ammonia compared to optical and IR dust emission
(Smith et al.\ 2004; Chesneau et al.\ 2005).  Our observations are
only sensitive to the brightest ammonia emission associated with the
warmest dust features near the star.  We highlight three main velocity
components in NH$_3$ (the numbers below are the numbered features in
Figs.\ 1 and 2):

1) The brightest ammonia emission is found at +56 km s$^{-1}$ in
   Figure 1 and centered $\sim$0$\farcs$2 E/SE of the star in Figure
   2, corresponding to the point where the back wall of the
   approaching SE polar lobe meets the equator in the tightly-pinched
   waist of the bipolar nebula.  The ammonia emission overlaps exactly
   with the brightest H$_2$ emission in velocity, and is therefore
   behind the continuum source (see Smith 2006).  This indicates that
   polyatomic molecules can survive very close to the star (within a
   few hundred AU) in dusty clumps seen in high resolution images
   (Chesneau et al.\ 2005; Smith et al.\ 2004).  The required
   shielding is evidently from a column of $\ga$10$^{23}$ cm$^{-2}$
   and $A_V\ga$4 mag (Smith 2006).  Chesneau et al.\ found unusual
   aluminum-rich dust composition in these blobs, which may be related
   to their molecular chemistry and the C and O depletion.  If this
   gas is at 140--250 K (see below), then the emission optical depth
   of this feature is 0.045 to 0.09.

2) The brightest H$_2$ emission is seen at --73 km s$^{-1}$ and
   1\arcsec\ NW of the star in Figure 1a.  It is the blushifted
   counterpart of component 1.  This blueshifted feature is not seen
   in NH$_3$ emission, but it is seen weakly in {\it absorption} at
   the same velocity.  Figure 2 shows that this molecular absorption
   coincides with the brightest equatorial dust condensations near the
   star (Chesneau et al.\ 2005).  The ammonia absorption also overlaps
   spatially with bright ionized blobs called the Weigelt knots (see
   Smith et al.\ 2004).  These knots are known to be in the equator on
   the near side of the star.  Thus, the ammonia molecules probably
   absorb the direct 12mm free-free emission from their own ionization
   fronts.  The strength of the absorption feature is about --3.5 mJy
   (Fig.\ 1b), which would correspond to an optical depth of about
   0.002 if it is absorbing the $\sim$2 Jy continuum source.  However,
   the actual optical depth depends on how much of the extended 12mm
   continuum is covered by the absorbing condensations, which is very
   uncertain at our spatial resolution.  In order for $\tau_{abs}$ in
   component 2 to match $\tau_{em}$ for component 1, the covering
   factor would be 4--8\%.

3) We also detect a third ammonia feature not seen in H$_2$ emission,
   located at --110 km s$^{-1}$ in Figure 1 and $\sim$1\arcsec\ NW of
   the star in Figure 2. Its velocity clearly does not match the wall
   of the NW polar lobe seen in H$_2$.  Instead, it probably
   originates from the so-called ``strontium filament'' (see Fig.\ 2a
   and Hartman et al.\ 2004).  Unusual low-ionization atomic emission
   lines like [Sr~{\sc ii}], [Ca~{\sc ii}], [Ni~{\sc ii}], etc., are
   seen to be localized at this same position and velocity (Davidson
   et al.\ 2001; Smith 2002a; Hartman et al.\ 2004).  The presence of
   NH$_3$ emission combined with the lack of infrared H$_2$ emission
   here may give important clues to the excitation of the ``strontium
   filament''.  In particular, it indicates that there is warm
   molecular gas in the equatorial plane that is shielded from the FUV
   radiation which pumps the Lyman-Werner bands of H$_2$.  The fact
   that the NH$_3$ emission in component 3 is found where the
   visual-wavelength emission lines from the ``strontium filament''
   seem to end is probably due to our line-of-sight extinction -- this
   coincides with the point where equatorial material closer to the
   star is blocked by dust in the edge of the SE polar lobe.  Figure 1
   may indicate a related component at positive velocities as well, at
   roughly +185 km s$^{-1}$.

The NH$_3$ data also show a possible absorption feature at --500 to
--550 km s$^{-1}$ along the line of sight to the star (not shown), but
it is difficult to judge the reality of this feature since it was near
the edge of our observed bandpass.  We might expect absorption at that
velocity from the approaching wall of the SE polar lobe of the
Homunculus (Gull et al. 2005; Smith 2002a, 2006).

%%%%%%%%%%%%%%%%%%%%%%%%%%%%%%%%%%%%%%%%%%%%%%%%%%%%%%%%%%%%%%%%
\section{DISCUSSION: THE RELEVANCE OF AMMONIA}

The Homunculus is the only LBV ejecta nebula known to contain
molecular gas\footnote{There is evidence for CO J=1-0 and 2-1 emission
associated with AG Car (Nota et al.\ 2002) and 2.3~$\micron$ bandhead
emission associated with HR Car (McGregor et al.\ 1988), but in both
cases the authors conclude that the CO emission arises in the
immediate circumstellar environment of the star, and not in the
detached nebulae resolved in optical emission-line images.}, and our
detection of NH$_3$ is the first detection of emission from a
polyatomic molecule in the ejecta around any LBV.  The detection of
ammonia raises interesting questions about how complex molecules can
form rapidly in the ejecta of a hot star.  Did these molecules grow
directly in the gas phase, or did they accumulate on grain surfaces
first?  Were they expelled from icy grain mantles by shocks?
Polyatomic molecules are harder to form than simple diatomic molecules
in an extreme irradiated environment, so these molecules likely formed
when the ejecta were very dense and self-shielding.  The major dust
formation episode happened $\sim$15 yr after the Great Eruption's 1843
peak, judging by the sharp drop in the visual brightness.  If
molecules formed at the same time, then scaling back from the
currently-observed density in the H$_2$ skin of the Homunculus (Smith
2006), the density and temperature in the ejecta at that time were
10$^{8.5}$--10$^9$ cm$^{-3}$ and 1000--1500 K.

While bright near-IR emission from H$_2$ has clearly been detected in
the Homunculus (Smith 2002a, 2006), CO has never been detected, either
in 2.3~$\micron$ bandhead emission (Smith 2002a) or in $^{12}$CO
J=2--1 (Cox \& Bronfman 1995).  The abscence of CO combined with a
clear detection of NH$_3$ provides a qualitative indication that the
molecular gas in the Homunculus is indeed nitrogen rich and has been
processed through the CNO cycle.  This is not necessarily surprising
in light of the other CNO-processed ejecta around $\eta$ Car (Davidson
et al.\ 1982; Dufour et al.\ 1997; Smith \& Morse 2004), but it is the
first actual observation that the Homunculus itself is also N-rich
because atomic abundances for the neutral Homunculus cannot be
obtained easily.

Calculating an ammonia abundance from our current data is frought with
uncertainty, which can be reduced with future observations of other
ammonia transitions.  For component 1 in Figure 1, we measure a total
flux integrated over the line of $\sim$1.08 Jy km s$^{-1}$, or a
beam-averaged brightness temperature of $T_B \simeq$ 11.3 K (with an
adopted beam filling factor of 1).  The column density of NH$_3$ (3,3)
is given by

\begin{displaymath}
N_{33} = \frac{8 \pi {\rm k} \nu^2}{A_{33} \ {\rm h} \ c^3} \ T_B \ \Delta\nu
\end{displaymath}

\noindent (e.g., Rohlfs \& Wilson 2004) where
$A_{33}$=2.56$\times$10$^{-7}$ s$^{-1}$ (Ho \& Townes 1983), and
$\Delta\nu$=1.99 MHz (25 km s$^{-1}$).  The observed intensity of
component 1 indicates N$_{33}\simeq$10$^{15}$ cm$^{-2}$.  With only
one transition, we don't know the gas temperature.  However, at $n_H >
10^6$ cm$^{-3}$ in the Homunculus walls (Smith 2006; Gull et al.\
2005), it is likely that the gas and dust temperatures are in
equilibrium, so we take the range of observed dust temperature of
140--250 K (Smith et al.\ 2003).  (We also neglect the possibility of
non-thermal amplification.)  If we adopt the expression for the
partition function given in equation A15 of Ungerechts et al.\ (1986),
we find a total ammonia column density of 10$^{15.7}$ cm$^{-2}$
assuming a temperature of 140 K, or 10$^{15.8}$ cm$^{-2}$ assuming 250
K.  The column density of H$_2$ as we look perpendicularly through a
wall of the Homunculus should be several $\times$10$^{22}$ cm$^{-2}$
(Smith 2006; Ferland et al.\ 2005).  Taken together, we therefore find
a rough estimate for the ammonia abundance of
N(NH$_3$)/N(H$_2$)=(1.6--2)$\times$10$^{-7}$.  This is almost a factor
of 10 higher than ammonia abundances derived recently in a similar way
for a sample of infrared dark clouds by Pillai et al.\ (2006), and our
low estimate is higher than the highest value in their sample. Because
of differences in beamsize and velocity range, the limits to CO
emission given by Cox \& Bronfman (1995) are not enough to place
meaningful constraints on the N(NH$_3$)/N(CO) abundance when compared
with our data; renewed attempts to detect CO in $\eta$ Car would be
useful.

Unfortunately, we cannot provide a quantitative estimate of the total
nitrogen abundance in the ejecta from ammonia observations alone, as
it depends upon grain chemistry, whether or not ammonia has been
liberated from icy grain mantles, and the ammonia survival probability
at various locations in the nebula.  Furthermore, most of the nitrogen
in the molecular zones is expected to be locked up in unobservable
N$_2$ molecules (Ferland et al.\ 2005).  It will be interesting if
future observations of additional transitions with higher sensitivity
can constrain the ammonia abundance as a function of position
throughout the Homunculus, and if future observations can detect
additional molecules that we might expect in nitrogen-rich molecular
ejecta.  For instance, N$_2$H$^+$ may be expected as part of the path
to NH$_3$.

\acknowledgements \scriptsize

We thank Bob Sault for assistance with the October 2004 observations,
and E.\ Keto for helpful discussions.  The July 2005 data were
obtained through director's discretionary time.  We also thank the
ATNF Distinguished Visitor Program, providing us with an opportunity
to collaborate on this project in Australia.  N.S.\ was supported
through grant HF-01166.01A from the Space Telescope Science Institute,
operated by the AURA, Inc., under NASA contract NAS~5-26555.

\end{document}